\documentclass[amsmath,amssymb,prb]{revtex4}
\usepackage{amsmath}
\usepackage{amssymb}
\usepackage{listings}
\usepackage{dcolumn}
\usepackage{bm}
\usepackage{graphicx}
\usepackage{dcolumn}
\usepackage{bm}
\usepackage{url}
\def\figno#1{Fig.~\ref{fig:#1}}
\def\ns{{\cal N}}
\def\fPeak{f_{{\rm peak}}}
\def\eqnn#1{Eq.~(\ref{eq:#1})}
\def\secno#1{section~\ref{sec:#1}}
\def\tobs{t_{{\rm o}}}

\def\tObservation{{{\cal  T}_{\rm obs}}}
\newcommand{\cs}{c}
\newcommand{\ts}{t_{\mathrm{s}}}

\newcommand{\freqs}{f_{\mathrm{s}}}
\newcommand{\unit}[1]{\,\mathrm{#1}}
\begin{document}
\title{Direct quantitative measurements of Doppler effects for sound
  sources with gravitational acceleration}
\author{
  Kenichiro Aoki$^1$\footnote{E--mail addresses:~{\tt
      ken@phys-h.keio.ac.jp,
    mitsui@hc.keio.ac.jp, yamamoto@koeki-u.ac.jp}.},
  Takahisa Mitsui$^1$
  and Yuki Yamamoto$^{1,2}$
}
\affiliation{
  $^1$Research and Education Center for Natural Sciences,  
  and 
  Dept. of Physics, Keio University, {\it
    4---1---1} Hiyoshi, Kouhoku--ku, Yokohama 223--8521, Japan\\
  $^2$Tohoku University of Community Service and Science, 
  Iimoriyama 3-5-1, Sakata 998-8580, Japan
}
  \date{\today }
\begin{abstract}
    We explain simple laboratory experiments for making quantitative
    measurements of the Doppler effect from sources with acceleration.
    We analyze the spectra and clarify the conditions for the Doppler
    effect to be experimentally measurable, which turn out to be
    non-trivial when acceleration is involved.  The experiments use
    sources with gravitational acceleration, in free fall and in
    motion as a pendulum, so that the results can be checked against
    fundamental physics principles.  The experiments can be easily set
    up from ``off the shelf'' components only.
    The experiments are suitable for a wide range of students,
    including undergraduates not majoring in science or engineering.
\end{abstract}
\maketitle
\section{Introduction}
\label{sec:intro}
The Doppler effect is a fundamental aspect of wave phenomena 
important in almost all areas of physics. The effect can be
experienced in everyday life from fast moving cars or trains passing
by and is also used as a practical method for measuring the velocity
of objects such as baseballs and automobiles. It is
also of crucial importance in astrophysics, where the receding or
approaching velocity of an object is determined using the Doppler
effect in light.  The results are used as a basis for such fundamental
aspects as Hubble's Law and the existence of black holes.
We believe that the Doppler effect, given its importance and
practicality, is a quite suitable theme as a basic physics experiment
for all types of students.

In this work, we explain concretely experiments for quantitatively
measuring the Doppler effect due to the motion of the wave source
with gravitational acceleration. 
Our purpose is to design experiments for undergraduate 
students including non-science majors. As such, we need an experiment
that can be easily set up and safely conducted in an ordinary
laboratory situation. Further, data acquisition needs to be robust and not
sensitive to the environment.  The students should be able to clearly
understand what is going on and the results should be quantitative.
Since we have a wide variety of students in mind, the analysis should
not necessitate ``higher'' mathematics and desirably, the setup should
use inexpensive components which are readily available.  As explained
below, we have been able to achieve all these goals.
Since practical considerations are of import, we provide the details
of the experiment, such as the parameters and the equipment used.
Due to its significance as a fundamental physics phenomenon, various
forms of Doppler effect experiments have been proposed and conducted
for some time: Doppler effect due to a wave source on a
toy car\cite{Barnes}, on an air track\cite{Warsh,Kosiewicz,Nerbun}, in
circular motion\cite{Gagne,Saba03,Spicklemire}, carried by a
student\cite{Cox} and in free fall\cite{Bensky,Torres} as well as the
effect due to the motion of the receiver\cite{Zhong} have been studied.
Demonstrations of Doppler effects from real cars have also been
performed\cite{Rossing,Saba01}.
However, the majority of the past experiments seem either rather
difficult to conduct, use ultrasonic frequencies, not quantitative or
require specialized equipment or circuitry.

We explain the basic principle and the setup of the experiments in
\secno{setup}.  Section \ref{sec:freefall} contains the details and
the results 
from Doppler effect measurements of a freely falling source. We
analyze the properties of Doppler effects from accelerating
sources in \secno{acc}.  In the process, we clarify some
non-trivial conditions and limitations in measuring Doppler effects.
We explain the Doppler effect measurements of a
pendulum source in \secno{pendulum}.

\section{Experimental principle and setup}
\label{sec:setup}
In this work, we directly measure the Doppler shift due to the motion
of the source, since this approach is most often used in various areas of
fundamental physics and in practical measurements.
The principle of the experiments is to measure the
velocity of a moving wave source using the Doppler effect relation,
$    {f/ f_s}=  {1/( 1-v/\cs)}$.
The measured frequency is $\freqs$ when the source is stationary and
$f$ when it is approaching the observer with velocity $v$.
$\cs$ is the speed of sound.  From the time dependence of $v$, we can
obtain the acceleration. While this principle can indeed be made to
work, the simplicity is somewhat deceptive, even in principle. To
measure a frequency at any given instant, a finite amount of time $T$
is necessary and the instantaneous velocity picture ceases to hold
for large accelerations or for long measurement times, as explained in
\secno{acc}.  ($T$ is different from the time of observation,
$\tObservation$, during which we make multiple measurements of the
frequency.)

While the idea is simple, one main obstacle is that $v$ is at most
few\,\% of $c$, due to practical considerations, including safety.
Consequently, to measure $v$ to 10\,\% accuracy, for
instance, we need 0.1\,\% level accuracy in the
measurement of $f$, which is not trivial.
A theoretical limitation exists that to obtain a frequency resolution
$\Delta f$, time $1/\Delta f$ is necessary for the measurement.  To
obtain the desired precision, we use a relatively high frequency sound
$\sim4\,$kHz and require $\Delta f\lesssim10\,$Hz. Consequently, the
amount of time required for the measurement is $T\gtrsim0.1$\,s.  A
free fall from the height of 2\,m takes $0.64\,$s, for instance, which
is not that much larger than $T$. 
Increasing $T$ to decrease  $\Delta f$ can conflict with
measuring Doppler effects from accelerating sources, as explained in
\secno{acc}.
$T=\ns/R$ so that $\Delta f=1/T=R/\ns$, where $R$ is the sampling rate
and $\ns$ is the number of samples. 
Any $R$ can be used as long as $R$ is higher than
twice the frequency of interest. 
$\ns$ is chosen
appropriately depending on the situation at hand.
Since $\Delta f$ is fixed by $T$, larger $\freqs$ can achieve higher
relative accuracy.  For the clarity of exposition to students and
since  generic components we utilize are not guaranteed to
function properly outside the audible range, we use an audible sound
source.
All the experiments below were conducted with $\freqs=3.520\,$kHz,
a clearly audible sound while not being too unpleasant.

In the experiments, we need a simple way to measure frequencies in a
short period of time with adequate accuracy.  For this we use a
microphone attached to a personal computer (PC) and a spectrum
analyzer software. While most microphones work, one with high
directivity facilitates the experiment.
For spectral analysis, we used the GNU Octave software\cite{Octave}
and an explicit example script for extracting  spectra from a sound
file is provided in Appendix~\ref{sec:octave}.
An important technical point is to find an object that emits a sound
at a distinct frequency steadily and a digital voice recorder is
suitable for this task. It has the additional advantage that the sound
can be synthesized on a PC, transferred to it, then played back.
Synthesizing such a sound file can be accomplished by using GNU
Octave\cite{Octave} or SoX\cite{sox}, for instance.
A generic recorder with sufficient sound volume can perform the task
and a means to directly transfer the sound from a PC is convenient.
We also tested various simple buzzers for this purpose, but none of
those tested could emit a sound with a steady enough frequency,
especially when in motion.
Instead of a digital voice recorder, in some cases, a speaker attached
to the same PC can be used to replay the synthesized sound, which can be
somewhat cheaper. However, in this case, enough volume needs to be
attainable
and we need to use a software that does not emit the sound as it is
recording, since it will otherwise interfere. Having an independent
playback device such as a digital recorder, we find, is versatile,
convenient and facilitates the experiment.
The components we used are an inexpensive PC, Eee~PC (ASUS,
$\sim\$300$), the microphone Audio Technica AT815b ($\sim\$200$) and
the voice recorder, Olympus DS-51 ($\sim\$100$). All these components
are generic and, in our experience, other similar components also
suffice.
\section{A free fall Doppler experiment}
\label{sec:freefall}
\begin{figure}[htbp]
    \centering
   \includegraphics[width=7.5cm,clip=true]{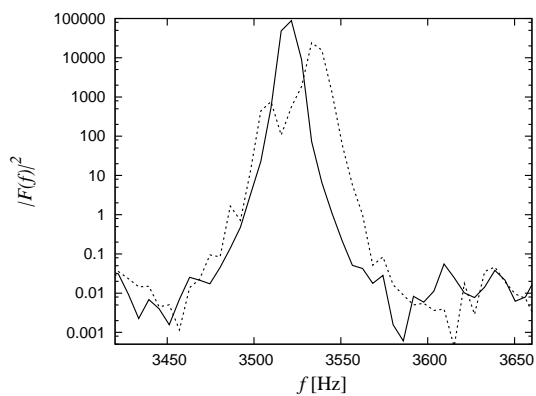}
   \caption{
     Examples of the measured sound spectrum, from the
     source at rest (solid) and when approaching the detector during
     free fall (dashed). The peak is Doppler shifted in the latter
     case.  The subdominant peak corresponds to the reflection off the
     ceiling and is Doppler shifted as a  receding source.
     The overall scale of $|F(f)|^2$ is arbitrary and Hann
     window is used.
}
    \label{fig:spectrum}
\end{figure}
\begin{figure}[htpb]
    \centering
    \begin{tabular}{cc}
        \includegraphics[width=8cm,clip=true]{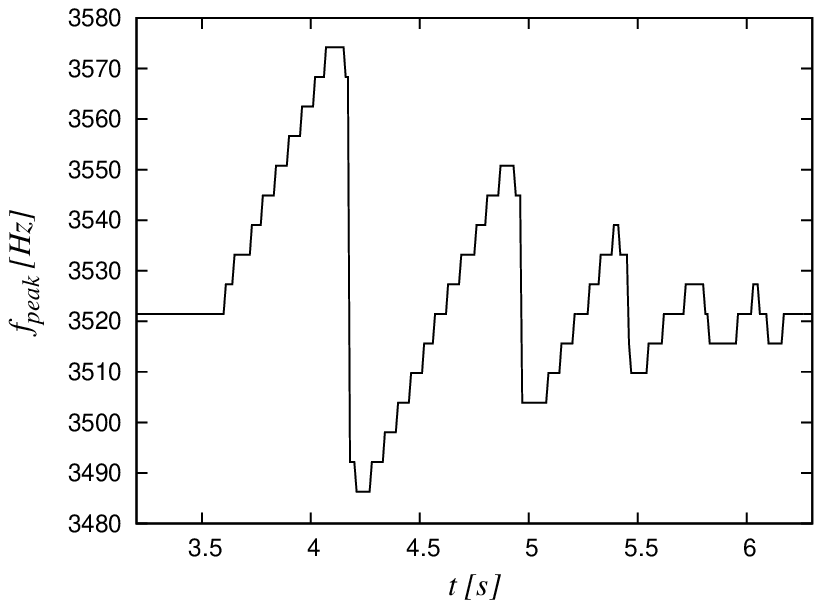}&
        \includegraphics[width=8cm,clip=true]{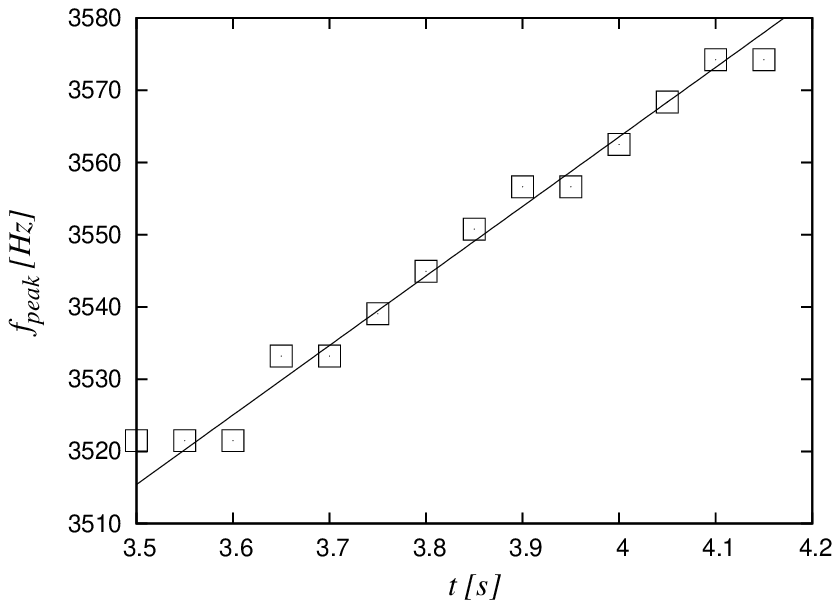}
    \end{tabular}
  \caption{Doppler shifts measured from a freely falling sound
    source.
    (Left) Measured peak frequencies with  $\Delta
    t=0.01$\,s. After 4.2\,s, the line is stretched to maximum
    and the source is pulled back up and falls again, which repeats.
    The frequency resolution $\Delta f=6$\,Hz is clearly visible. 
    (Right) An example of an analysis performed in a student
    experiment. 
    The  same data as the previous plot, in the region $t=3.5$ to
    $t=4.2$ with $\Delta t=0.05$\,s. 
    A linear fit to the Doppler shifted frequencies is shown and
    the slope corresponds    to $g=9.4\pm0.7\unit{m/s^2}$  (ambient
    temperature 17$^\circ$C).   
  }
    \label{fig:vt}
\end{figure}

In this experiment, we let a sound source fall freely towards a
microphone and measure the time dependence of the frequency.
The experiment is similar in
spirit to some previous experiments\cite{Bensky,Torres}, though more
quantitative.
To obtain the experimental results, we start recording on the PC and
let the voice recorder drop (attached to a line to avoid it dropping all
the way). The spectrum is analyzed at every $\Delta t$ and the peak
frequencies are obtained.
Example spectra from a stationary and a freely falling source
approaching the observer are shown in \figno{spectrum}. 
Here, we used
$\ns=2^{15}, R=192$\,kHz taking into
consideration the constraints of \secno{acc}. Consequently,
$\Delta f=6\,$Hz with an error $\Delta v=0.6$\,m/s in $v$. 
An example of the time dependence of $f$ is shown in
\figno{vt}~(left), where we can see the Doppler effect from a source
falling repeatedly, after being pulled back up by the line.

In the student experiments, we let the students read off the peak
spectrum frequencies with $\Delta t=0.05$\,s or $0.1$\,s.  Then the
students plot these values against time and obtain the slope
(\figno{vt}~(right)). Using $f=\freqs/(1-at/c)\simeq \freqs(1+at/c)$
($t=0$ shifted here), they obtain the gravitational acceleration, $g$,
from the slope.  The students often discover the ``bouncing effect''
in \figno{vt}~(left) during the analysis and it is educational for
them to realize what they are observing.
By using free fall for this experiment, the students can also clearly observe
the constant acceleration due to gravity, which can  contribute to
the understanding of the gravitational force as well as the Doppler
effect.  
$g$ is measured within 10\,\% accuracy, enough to convince the student
of the mechanism and the effectiveness of the Doppler shift
measurements. The point here is confirm the understanding of the
Doppler effect by obtaining $g$ and its precision of $g$ is not the
main objective.
 Indeed, if measuring $g$ were the primary objective, we can
do so to $0.1\,\%$ accuracy using a simple pendulum.
\section{Doppler effect from accelerating sources}
\label{sec:acc}
In this section, we derive and explain the spectral properties of the
Doppler effect from accelerating sources briefly.
The practical
problem is the following: Consider an accelerating monotonic source.
We might expect the measured spectrum to have a peak Doppler shifted
by the instantaneous velocity of the source, but  a finite time is needed
to make the measurement so no ``instantaneous'' velocity is measured.
We provide quantitative criteria for observing a simple Doppler
shifted spectrum and clarify what exactly can be measured.

When a source approaching the observer with a constant acceleration,
$a$, emits a signal at time $\ts$, it is received by the observer at
time $t$, and their relations are $ \cs(t-\ts)=L-{a\ts^2/2}$.
Here, without loss in generality, the initial distance between the source
and the observer is $L$ and the source is stationary when $\ts=0$.
If the signal has a definite frequency $\freqs$, it can be reexpressed
in terms of the observer time as
\begin{equation}
  \sin(2\pi \freqs \ts)  =
  \sin \left[  2\pi \freqs \left(t-{L\over\cs}\right)
    \left(1+{a\over2\cs}\left(t-{L\over\cs}\right)
    \right)\right]\qquad.
\end{equation}
Here, we assumed the source velocity $v$ is such that 
$v\ll\cs$ and have kept the leading order terms.
The signal is measured by the observer from time $\tobs$ to $\tobs+T$ and
its power spectrum can be obtained from the Fourier transformed
signal,
\begin{eqnarray}
  F(f)
  &=&
  \int_{\tobs-L/\cs}^{\tobs+T-L/\cs}dt\;
  \sin\left[2\pi \freqs\left(1+\frac{\gamma\freqs t}{2}\right)t\right]
  e^{i2\pi ft}
  \frac{1}{2if\sqrt{2\gamma}}
  \left\{
    \tilde{C}(f)\cos{\pi\over\gamma}\left(1+{f\over\freqs}\right)^{2}
    +\tilde{S}(f)\sin{\pi\over\gamma}\left(1+{f\over\freqs}\right)^{2}
      \right.
    \nonumber\\ &&\qquad
    -\tilde{C}(-f)\cos{\pi\over\gamma}\left(1-{f\over\freqs}\right)^{2}
    -\tilde{S}(-f)\sin{\pi\over\gamma}\left(1-{f\over\freqs}\right)^{2}
    +i\left[
      -\tilde{C}(f)\sin{\pi\over\gamma}\left(1+{f\over\freqs}\right)^{2}
+\tilde{S}(f)\cos{\pi\over\gamma}\left(1+{f\over\freqs}\right)^{2}\right.
    \nonumber\\ &&\qquad
    \left.
\left.-\tilde{C}(-f)\sin{\pi\over\gamma}\left(1-{f\over\freqs}\right)^{2}
      +\tilde{S}(-f)\cos{\pi\over\gamma}\left(1-{f\over\freqs}\right)^{2}
    \right]
  \right\}\quad.
  \nonumber
\end{eqnarray}
Here we defined a dimensionless parameter,  $\gamma\equiv
a/(\cs\freqs)$. 
$\gamma$ is the relative change in the velocity in one period of the
sound oscillation so that $\gamma\ll1$.
\def\yZero{y_{{\rm o}}}
We also defined $
  \tilde{C}(\pm f)
  \equiv
  C\left(x_\pm,\yZero\right)
  ,
  \tilde{S}(\pm f)
  \equiv
  S\left(x_\pm,\yZero\right)$ where 
  $  x_\pm\equiv\sqrt{2\gamma}\freqs\left(\tobs-{L /  c}\right)
      +\sqrt{2 / \gamma}\left(1\pm{f / \freqs}\right),\quad
    \yZero\equiv\sqrt{2\gamma}\freqs T
$.
These functions are related to the Fresnel integrals and their
necessary properties are explained in Appendix~\ref{sec:Fresnel}.
The  power spectrum is
\def\phiZero{\varphi_{{\rm o}}}
\begin{eqnarray}
    &&|F(f)|^{2}  =  \frac{1}{8\gamma\freqs^2 }
    \left(G_0+G_1+G_2\right), \
    G_0  \equiv    \tilde{C}(-f)^{2}     +\tilde{S}(-f)^{2}\\\nonumber
    &&G_1 \equiv 
    -2\left[\tilde{C}(f)\tilde{C}(-f)-\tilde{S}(f)\tilde{S}(-f)\right]
    \cos{\pi\over2}\phiZero
    -2\left[\tilde{C}(f)\tilde{S}(-f)+\tilde{S}(f)\tilde{C}(-f)\right]
    \sin{\pi\over2}\phiZero
    G_2 \equiv      \tilde{C}(f)^{2}     +\tilde{S}(f)^{2}\nonumber
    \quad,
  \label{eqn:F012}
\end{eqnarray}
where we defined $\phiZero\equiv4(1+f^2/\freqs^2)/\gamma$. 
The spectrum is involved, but we first note that when $f\simeq\freqs$,
$x_+={\cal O}\left(\gamma^{-1/2}\right)\gg1$.
Therefore, using the asymptotics of the Fresnel functions in
\eqnn{FresnelAsymptotic}, ${G_1/ G_0}={\cal
  O}\left(\gamma^{1/2}\right),{G_2/ G_0}={\cal O}\left(\gamma\right)$
and both are much smaller than 1, when $f\sim \freqs$.
We now analyze the dominant term $G_0$ and then show this suffices.  As
explained in Appendix~\ref{sec:Fresnel}, $G_0$ has a well defined peak
when
\begin{equation}
    \label{eq:peakCondition}
    \yZero=\sqrt{2\gamma}\freqs T< C_0    
    \quad\Leftrightarrow\quad
    \freqs T^2<{c\over 2a}C_0^2
    \quad,
\end{equation}
where $C_0=3.073490$, numerically.  $\yZero$ is the product of (the
square root of) the acceleration relative to the sound velocity and
$T$, both measured in the time scale $1/\freqs$.  Heuristically, this
condition is reasonable; $v$ is changing so that a
simple peak does not exist when $a$ is too large or 
$T$ is too long.  However, since both $\gamma$ and
$\freqs T$ are dimensionless, it is a priori not clear why this
particular combination is the relevant criteria and this concrete
computation enabled us to clarify the point.  The condition puts an
upper bound on $T$ which competes with the precision $\Delta f=1/T$,
so that judicious balancing is needed.
In all our experiments,  $T$ was chosen taking
these considerations into account. 
The peak, when it exists, is at
$ {\fPeak /  \freqs}-1={a /  \cs}  \left(\tobs-{L / \cs}+{T / 2}\right)$.
This Doppler shift corresponds to $v$ for which the signal is received
at the central time of the measurement, as expected. When
$\sqrt{2\gamma}\freqs T>C_0$, the spectrum has multiple peaks
and the determination of the ``peak'' becomes ambiguous.

We now analyze the effect of the subleading terms, $G_{1,2}$, to
$\fPeak$.  In general, when a function $g_0(f)$ is peaked at $f_0$,
the peak for $g_0(f)+\epsilon g_1(f)$ occurs at $f_0-\epsilon
g_1'(f_0)/g_0''(f_0)$, to leading order in $\epsilon$.  Therefore, we
need to know the size of $f\,dG_{1,2}(f)/df$.  Their individual terms are
${\cal O}(\gamma^{-1/2})$ which can give rise to shifts comparable to
the Doppler shift, so that a finer estimate of these terms is
necessary.  After some non-trivial computation, we obtain
\begin{eqnarray}
    \freqs{dG_1(f)\over df}\biggr|_{f=\fPeak} 
    &=& 
    -2\sqrt{2\over\gamma}\tilde C(-f)\biggl\{
      \cos{\pi\over2}\left[\left(x_++\yZero\right)^2
          -\phiZero\right]
      -\cos{\pi\over2}\left[x_+^2
          -\phiZero\right]
         \nonumber\\     &&\qquad
        +\pi(x_+-x_-)
    \left[\tilde S(f)\cos{\pi\over2}\phiZero 
      -\tilde
      C(f)\sin{\pi\over2}\phiZero\right]\biggr\}\biggr|_{f=\fPeak}
    \nonumber\\
    &&+2\sqrt{2\over\gamma}\tilde S(-f)\biggl\{
      \sin{\pi\over2}\left[\left(x_++\yZero\right)^2
          -\phiZero\right]
      -\sin{\pi\over2}\left[x_+^2
          -\phiZero \right]\\
    &&\qquad-\pi(x_+-x_-)
    \left[\tilde C(f)\cos{\pi\over2}\phiZero 
      +\tilde
      S(f)\sin{\pi\over2}\phiZero\right]\biggr\}\biggr|_{f=\fPeak}
    \nonumber\\
    \label{eq:dg2}
    \freqs{dG_2(f)\over df} &= &
    2\sqrt{2\over\gamma}\biggl\{\tilde C(f)\left[
      \cos{\pi\over2}\left(x_++\yZero\right)^2
      -\cos{\pi\over2}x_+^2\right]
        +\tilde S(f)\left[
            \sin{\pi\over2}\left(x_++\yZero\right)^2
            -\sin{\pi\over2}x_+^2\right]
          \biggr\}.\nonumber
    \label{eq:dg1}
\end{eqnarray}
The leading order terms in $\gamma$ cancel 
  and
$    \freqs{dG_1(\fPeak) /  df} = 
    {\cal O}\left(\yZero\right),
    \freqs{dG_2(\fPeak) /  df} = 
    {\cal O}\left(\gamma^{1/2}\right)$, 
using the asymptotic behaviors
\eqnn{FresnelAsymptotic}.
Consequently, the contributions from $G_{1,2}$ to the relative shift in
$\fPeak$ are ${\cal O}(\gamma\yZero)$, ${\cal
  O}(\gamma^{3/2})$, respectively. Since the relative shift in the
frequency due to the Doppler effect is
${\fPeak/\freqs}-1\geq\sqrt{\gamma/2}\,\yZero/2$, the relative error
in the shift due to $G_{1,2}$ is at most ${\cal O} (\gamma^{1/2})$.
We note that $\yZero$ is at most of order one from
\eqnn{peakCondition} and typically not much smaller than 1 due to the
required frequency resolution. 

\begin{figure}[htbp]
    \centering
    \begin{tabular}{cc}
        \includegraphics[width=8cm]{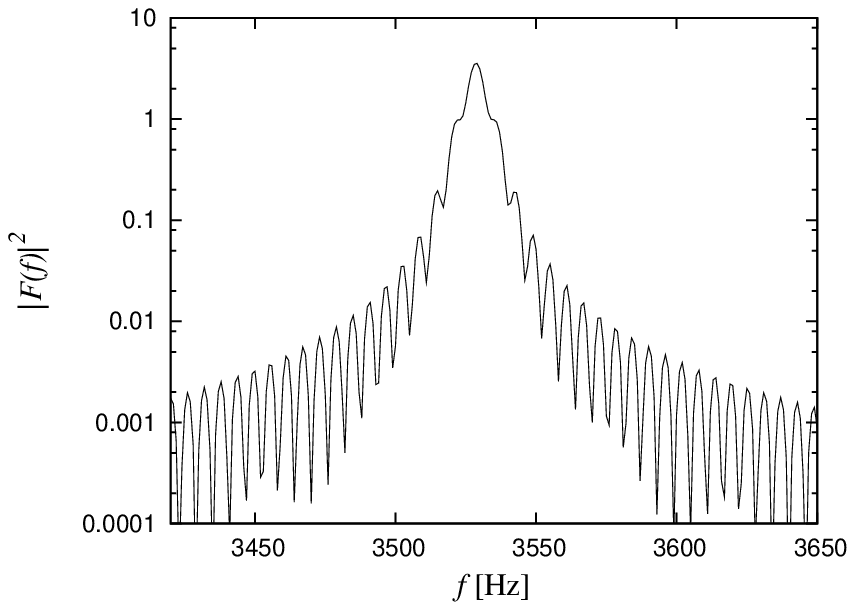}&
        \includegraphics[width=8cm]{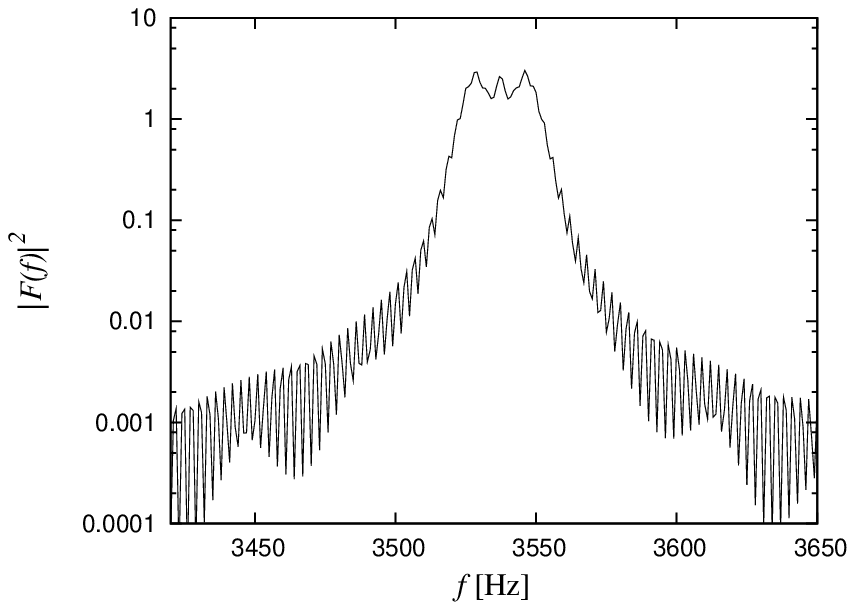}\\
        \includegraphics[width=8cm]{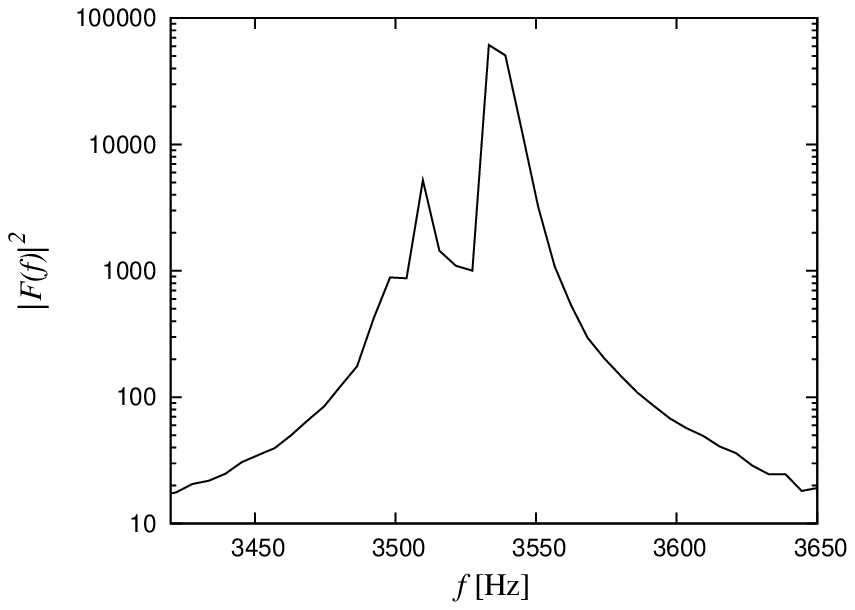}&
        \includegraphics[width=8cm]{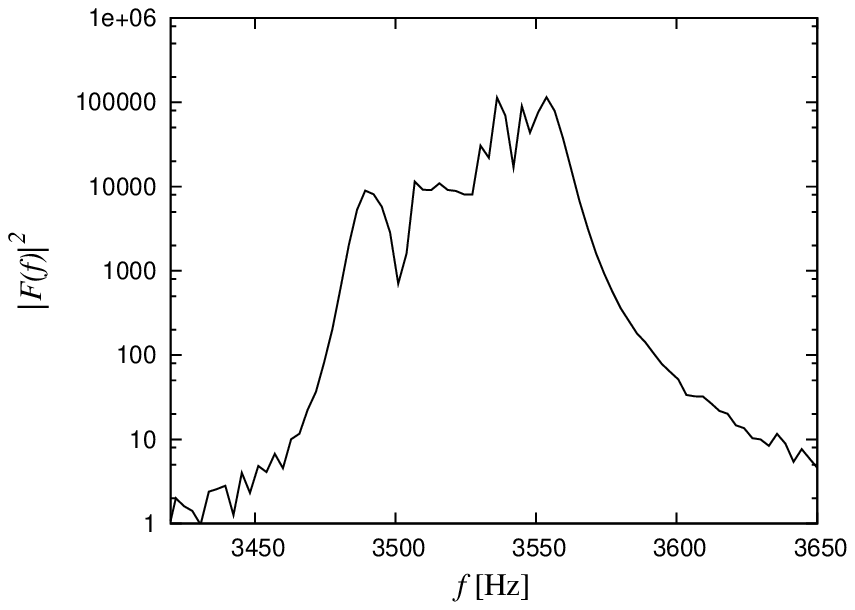}
    \end{tabular}
    \caption{(Left) The spectra for $R=192\unit{kHz},\ns=2^{15},
      \freqs T^2=102.5\,\unit{s}$, the simulated spectrum (top) and
      the experimental result (bottom). (Right) The spectra
      for $R=192\unit{kHz},
      \ns=2^{16}, \freqs T^{2}=410.1\,\unit{s}$, the simulated
      spectrum (top) and the experimental result (bottom).
      The subdominant peak (left bottom) is due to reflection,
      similarly to \figno{spectrum}.
      In all cases, no window function is used.  }
    \label{fig:multiPeaks}
\end{figure}

We now illustrate these points in the free fall experiment in
\secno{freefall}.
The condition for the existence of an unambiguous peak,
\eqnn{peakCondition}, corresponds to $\freqs T^{2}<166\,\unit{s}$. In
\figno{multiPeaks}, we compare the cases when this condition is and
is not satisfied both in simulations and in the analyses of the same
experimental data.
When condition \eqnn{peakCondition} is
satisfied, a distinct peak exists in the spectrum. However, when the
condition is not satisfied, a simple peak no longer exists even in
theory and it is impossible to ascertain the peak frequency in the
experimental situation.
Condition \eqnn{peakCondition} necessarily depends on the
acceleration and for small acceleration or for constant speeds, we can
use longer measurement times.  For instance, if we use
$\ns=2^{17}=131072$ (which is {\it not} appropriate for measuring
$g$), the error is $\Delta f=1\,$Hz, corresponding to $\Delta
v=0.1\,$m/s. Errors $\Delta f,\Delta v$ are inversely proportional to
$\ns, T$. However, due to \eqnn{peakCondition}, $\Delta
f,\Delta v$ can be decreased only as $\sqrt{a}$.

Let us summarize the situation: When measuring constant velocity, we
can reduce the error in $v$ by using longer $T$. However, there will
usually be practical limitations, such as one  can not let the
object move at constant velocity for an arbitrary amount of time nor
can we pick up the sound for a long time. When $a\not=0$, $T$ needs to
be small enough to satisfy \eqnn{peakCondition} for the Doppler shift
to be unambiguously observable.  Further, if we want to measure $a$
during time $\tObservation$, we have an additional requirement that
the Doppler shift needs to be larger than the frequency resolution.
Letting $\eta(<1)$ be the relative precision with which we measure
$a$, ${a\tObservation / c}>{1 / \eta C_0}\sqrt{2a / c\freqs} $,
equivalent to ${\tObservation / T}>{2 (\eta C_0\yZero)^{-1}}$.
For larger $a$, $\tObservation$ can be smaller, yet the attained
velocity difference is larger.
\section{Doppler effect measurements of a pendulum source}
\label{sec:pendulum}
\begin{figure}[htbp]
    \centering
    \includegraphics[width=7truecm]{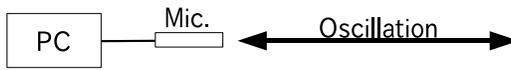}
    \caption{A top down view of the setup for a pendulum Doppler effect
      experiment.} 
    \label{fig:pendulumSetup}
\end{figure}
An experiment which provides more challenge in its analysis is the
measurement of Doppler effects from a source used as a pendulum. We
used the voice recorder strung by a line to a high beam as the
pendulum and the configuration of the experiment is shown in
\figno{pendulumSetup}. The strategy follows that used in the previous
sections, exemplifying the ease with which we can adapt our methods
to various Doppler shift measurements.
The measurement is easier if relatively large velocities are attained.
Consequently, we would like to a pendulum with a long length, $\ell$,
and a large initial angle.  Using typical values for a laboratory
situation, $\ell=2\,$m, $\theta_{\rm init}=\pi/4$, the maximum
velocity is $3.5\,$m/s which is quite measurable, as seen in the
previous sections.
The experiment is not as simple as it may seem for the following
reasons. First, if we use a large initial angle, the harmonic
approximation to the oscillation is no longer quite valid.
Furthermore, larger oscillations tend to die down rapidly. Smaller
oscillations can be observed for a longer period of time, yet
the velocities involved are smaller and the analysis becomes
technically more delicate.

\begin{figure}[htbp]
    \centering
    \begin{tabular}{cc}
        \includegraphics[width=8cm,clip=true]{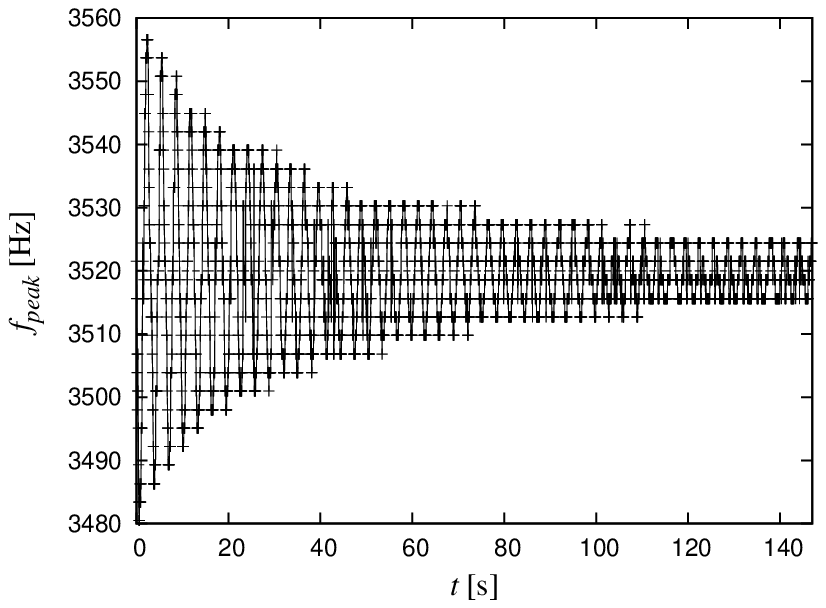}&
        \includegraphics[width=8cm,clip=true]{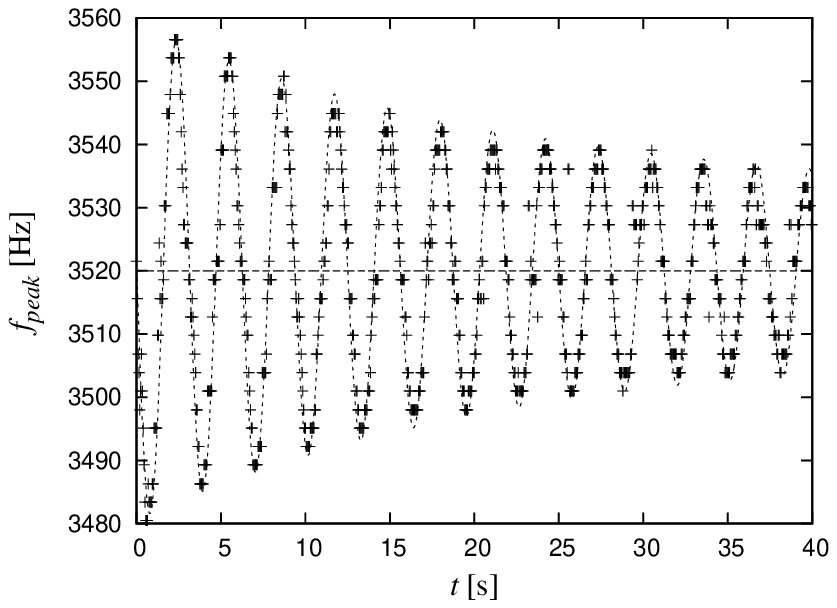}\\
        \includegraphics[width=8cm,clip=true]{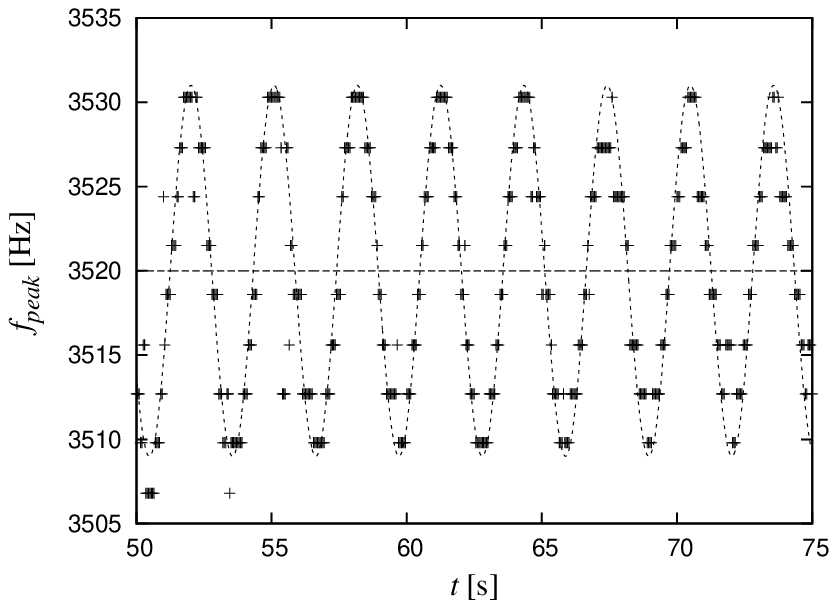}&
        \includegraphics[width=8cm,clip=true]{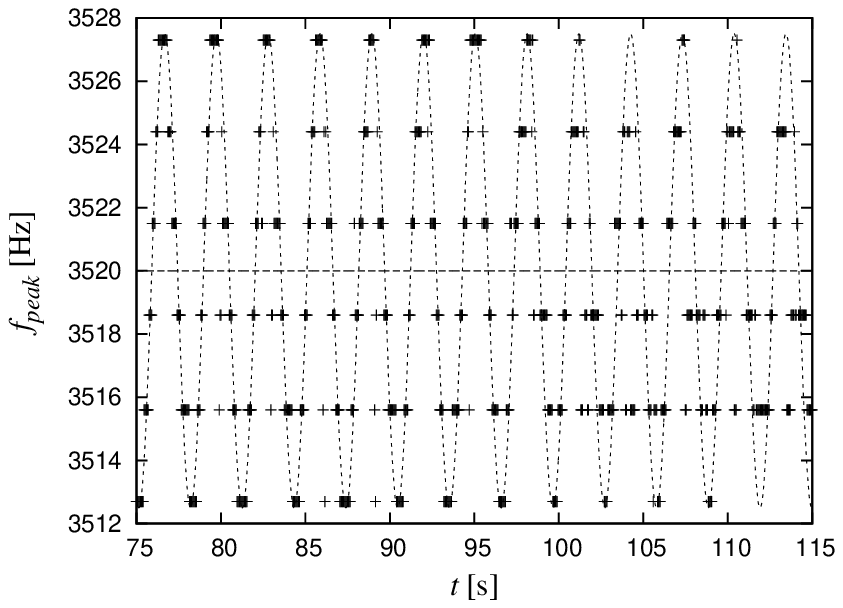}
    \end{tabular}
   \caption{(Top left) Peak frequencies measured from a sound source
     swinging as a pendulum. (Top right) Peak frequencies and a
     function of the form 
     $(a_0+b_0e^{-c_0t})\sin\omega_0(t-t_0)$ with $ a_0=11.5\unit{Hz},
     b_0=27.9\unit{Hz},c_0=0.045\unit{s^{-1}},\omega_0 =
     2.0125\unit{Hz},t_0=-1.54\unit{s}$, which describes the data in
     this region well.  The maximum velocity is 3.9\,m/s.
     (Bottom left, right) Peak frequencies and a simple
     function, $a_1\sin\omega_1(t-t_1)$ with $(a_1,\omega_1,t_1)=
     (11\unit{Hz},2.04\unit{Hz},1.97\unit{s}),
     (7.5\unit{Hz},2.05\unit{Hz},2.34\unit{s})$, respectively. The
     maximum velocity is $1.2,0.8$\,m/s, respectively. A simple
     harmonic oscillation can be seen to describe the data in these
     time intervals well.  
     $\freqs=3.52\,$kHz is also indicated in the plots.  }
   \label{fig:pendulum}
\end{figure}

Several approaches to analyzing this experiment are apparent.  One 
simple approach is to take various time periods of the oscillation
and compare them to theoretical predictions, as shown in
\figno{pendulum}.
In these examples, $\ell=2.32\pm0.01\,$m, the ambient temperature is
23$^\circ$C.  A lower sampling rate $R=48\,$kHz was used to reduce the data
size and $\ns=2^{14}$ was used to satisfy  \eqnn{peakCondition}. 
The observed oscillation frequency agrees with the standard formula,
$\sqrt{g/l}/(2\pi )$.  We can easily obtain $g$ to 1\,\% accuracy or
less and $g=9.8\pm0.1\unit{m/s^2}$ in this example.  If the objective
is solely to measure $g$, there is no distinct advantage over timing
measurements of a simple pendulum. For our purposes, the pendulum
source provides a variety of Doppler effects that can be
quantitatively measured, including approaching, receding and
accelerating sources all in one experiment, that can be checked
against basic physics. Another advantage is that
the time dependence of $v$ can be obtained more globally than that in
simple timing measurements.
A precise global description of the data, which is not attempted
here, needs to include the modeling of the decay in the oscillations
and the anharmonic effects in the pendulum. As a Doppler effect
measurement, we are quite satisfied that the oscillations can be seen
clearly and agree so well with the theory.  There are a number of
aspects to this experiment and more intricate analysis can be
performed if desired.  For instance, the change in the period due to
anharmonicity can be observed, as can be seen in the results in
\figno{pendulum}.  This kind of experiment might be appropriate for
more advanced students or for student projects.
\section{Discussion}
\label{sec:disc}
Considering the prevalence and the importance of the Doppler effect in
basic sciences, along with its practical usefulness, we believe that
easily realizable experiments in which the students can quantitatively
measure the Doppler effect are highly desirable.  In this work, we
explained two such experiments, which are suitable for
undergraduate students majoring in any field and require only generic,
off the shelf components. 
In the experiments, the source undergoes gravitational acceleration so
that the students can check the measured results against fundamental
physics principles.
The experiments are not difficult for
students to perform and one can visibly confirm the induced shift in
the peak of the audible sound. The setup we explained, as seen above,
is quite flexible and the experiment can be conducted with the sound
source attached to various objects in motion.  It might be interesting
to measure Doppler shifts of thrown objects, swung bats/rackets, and
so on, using this method.

In all our experiments, we picked out the highest peaks in the
spectrum and no attempt was made to identify the ``correct'' peak, yet
the Doppler effect could be measured quantitatively. The main reason
for the non-ideal spectra and the occurrence of multiple peaks  is the
existence of reflections which inevitable occur in ordinary student
laboratories.  In our experience, objects that are approaching and
therefore have higher peak frequencies tend to be easier objects for
measurement, since the interference effects do not affect the
highest frequency peak substantially. Our objective is to design
student experiments which can be performed in ordinary laboratory
situations and for such purposes, we designed the experiments with the
data extraction that is robust and clear cut.

The Doppler effect is a phenomenon that can be experienced in everyday
life and is not hard for students to understand.
One possible concern might be that students, particularly those not
majoring in science or engineering, find spectral methods hard to
comprehend. However, they are exposed to music and pitch in
everyday life and most students understand the concept of frequency
and harmonics, though not necessarily fully conscious of their
spectral context. Therefore, the shift in the peak frequency caused by
the Doppler effect is not a difficult concept for students to grasp.
We believe, rather, that experiments with sounds allow students to
pick up the idea of the spectrum and Fourier analysis intuitively and
serves as an excellent introduction to spectral methods.  In our
experiments, we combine the spectral analysis of sounds from musical
instruments with the free fall Doppler experiment in \secno{freefall},
in order for the students to first gain a good grasp of the relation
between sound, harmonics and the frequency spectrum. Undergraduate
students majoring in humanities and social sciences at Keio University
currently perform these experiments in class smoothly 
and we are quite satisfied with the outcome.
\appendix
\section{Extracting spectra from a sound file}
\label{sec:octave}
We provide an example script for GNU Octave (for Windows, Linux and
other Unix variants) that extracts spectra from a sound file at
predetermined time slices.
This example script plots the spectrum at times 
$0,\Delta t,2\Delta t,\cdots$, and displays the time (at the beginning
of each measurement) and the peak frequency on the terminal, 
given a sound file. It can be modified to look
at a particular region in time and window functions can also be
incorporated.
\begin{verbatim}
#!/usr/bin/octave -qf
[data,R]=wavread('freefall.wav'); # data, sampling rate from file
dt=0.05;                          # time between data points
N=2^15;                           # number of samples
fmin=3400; fmax=3700;             # freq. plot range
df=R/N;                           # frequency resolution
range=round((fmin/df):(fmax/df));
freq=(range-1)*df;                # x axis
printf("# t[s]\tfPeak[Hz]\n");
for i=round((0:floor(length(data)-N)/R/dt)*R*dt);
  spectrum=power(abs(fft(data(i+1:i+N))),2);	
  [specMax,jf]=max(spectrum);     # peak frequency
  printf("%.5g\t%.5g\n",i/R,(jf-1)*df); # time and peak frequency
  semilogy(freq,spectrum(range)); # spectrum, semilog plot
  pause;
endfor
\end{verbatim}
\section{Some properties related to Fresnel integrals}
\label{sec:Fresnel}
\begin{figure}[htbp]
    \centering
    \begin{tabular}{cc}
        \includegraphics[width=8cm]{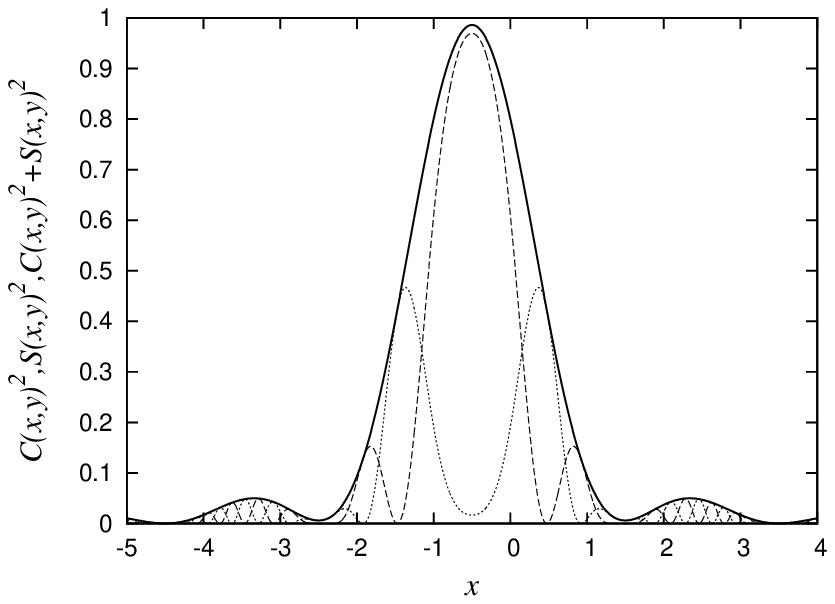}&
        \includegraphics[width=8cm]{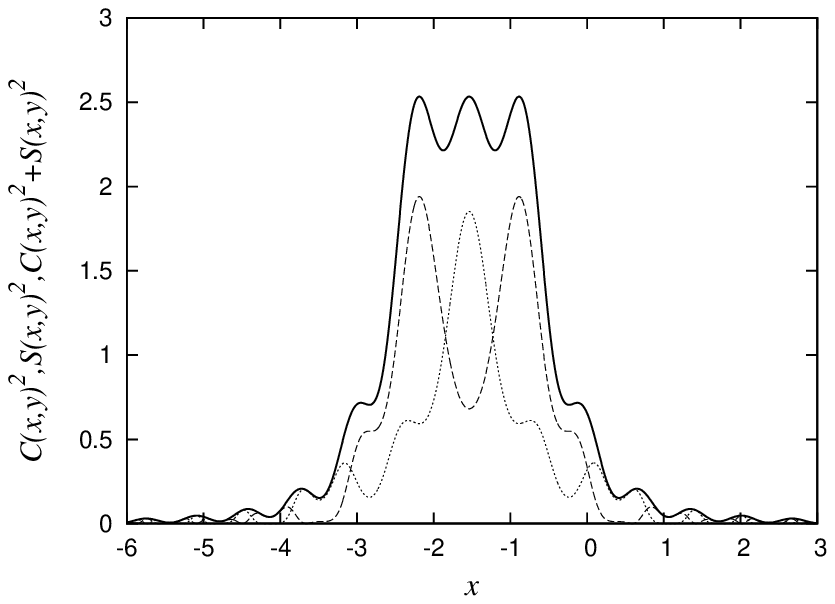}\\
        \includegraphics[width=8cm]{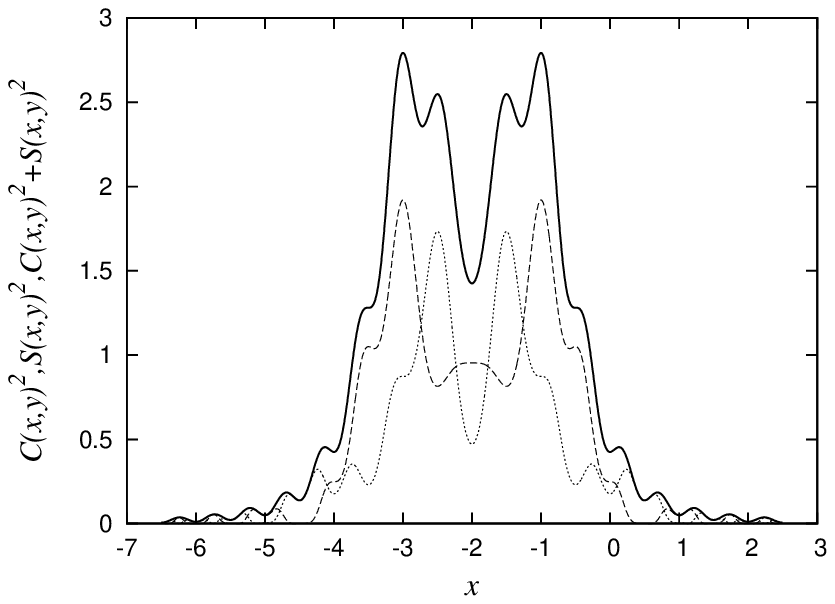}&
    \end{tabular}
    \caption{$C(x,y)^2$ (dashes), $S(x,y)^2$ (dots)  and 
      $C(x,y)^2+S(x,y)^2$ (solid) for $y=1$
      (left top), $y=3.0735$ (right top) and $y=4$ (left bottom).  
      The maximum at $x=-y/2$ disappears for $y>C_0$.  }
    \label{fig:cs2}
\end{figure}

We define the following functions
\begin{equation}
  C(x,y)\equiv C(x+y)-C(x)\quad,\quad
  S(x,y)\equiv S(x+y)-S(x)\quad.
\end{equation}
Here $C(x),S(x)$ are the Fresnel functions\cite{Fresnel}, 
\begin{equation}
    \label{eq:Fresnel}
  C(x)=\int_0^{x}dt\;\cos\left(\frac{\pi}{2}t^{2}\right)\quad,\quad
  S(x)=\int_0^{x}dt\;\sin\left(\frac{\pi}{2}t^{2}\right)\quad .
\end{equation}
We need the asymptotic behavior of the Fresnel functions which are, to
leading order\cite{Fresnel},
\begin{equation}
    \label{eq:FresnelAsymptotic}
    C(x) = {1\over2}+{\sin\left(\pi x^2/2\right)\over\pi x} 
      - {\cos\left(\pi x^2/2\right)\over\pi^2 x^3} + {\cal
        O}\left(x^{-5}\right),\ 
      S(x) = {1\over2}-{\cos\left(\pi x^2/2\right)\over\pi x} 
      - {\sin\left(\pi x^2/2\right)\over\pi^2 x^3} 
      + {\cal O}\left(x^{-5}\right).
\end{equation}
It is straightforward to find that the extrema of $C(x,y)$  are at 
\begin{equation}
  x=-\frac{y}{2}+\frac{2n}{y}, \ 
  \frac{-y\pm\sqrt{8m-y^{2}}}{2}\qquad(m\geq y^2/8)
\end{equation}
and similarly for $S(x,y)$, 
\begin{equation}
  x=-\frac{y}{2}+\frac{2n}{y},\ 
  \frac{-y\pm\sqrt{4(2m-1)-y^{2}}}{2}\qquad(m\geq y^2/8+1/2)\quad.
\end{equation}
Here, $m,n$ are integers.  The function $C(x,y)^2+S(x,y)^2$ has a
maximum at $x=-y/2$ when $y<C_0=3.073490$ as seen in \figno{cs2},
where the value of $C_0$ was determined numerically.


\begin{thebibliography}{99}
  \bibitem{Barnes}
    G. Barnes, ``A Doppler experiment'', Am. J. Phys. 42,
    905-909 (1974)
  \bibitem{Warsh}
    K. L. Warsh, R. T. Spires, and J. B. Hofmann, Jr., ``Doppler Radar
    for Air-Track Velocity Measurements'', Am. J. Phys., 35, 159-160
    (1967)
  \bibitem{Kosiewicz}
    R.M. Kosiewicz, ``The Doppler Effect and Lissajous Figures Using a
    Linear Air Track'', Am. J. Phys. 39, 229 (1971)
  \bibitem{Nerbun}
    R. C. Nerbun, Jr. and R. A. Leskovec, ``Quantitative
    measurement of the Doppler shift at an ultrasonic frequency'', Am.
    J. Phys. 44, 879--881 (1976)
  \bibitem{Gagne}
    R. Gagne, ``Determining the speed of sound using the
    Doppler effect'', Phys. Teach. 34, 126--127 (1996)
  \bibitem{Saba03}M. M. F. Saba and R. A. da S. Rosa, ``The Doppler
    effect of a sound source moving in a circle'', Phys. Teach. 41,
    89--91 (2003).
  \bibitem{Spicklemire}
    S.J. Spicklemire, M.A. Coffaro, ``The treatment of reflections in
    a Doppler measurement using the method of images'', Am. J. Phys.
    74, 40--42 (2006)
  \bibitem{Cox}
    A. J. Cox and Joel J. Peavy, ``Quantitative measurements of the
    acoustic Doppler effect using a walking speed source,'' Am.  J.
    Phys. 66, 1123--1124 (1998)
  \bibitem{Bensky}T.J. Bensky, S.E. Frey, 
    ``Computer sound card assisted measurements of the acoustic
    Doppler effect for accelerated and unaccelerated sound sources'',
    Am. J. Phys.  69, 1231-1236 (2001)
  \bibitem{Torres} S.M. Torres, W.J.~Gonz\'alez-Espada,
    ``Calculating g from Acoustic Doppler Data'', 
    Phys. Teach. 44, 536 (2006)
  \bibitem{Zhong} 
    A.~Zhong, ``An acoustic Doppler shift experiment with the
    signal-receiving relay'', Am. J.  Phys. 57, 49-50 (1989)
  \bibitem{Rossing}
    T.D.~Rossing, ``The Doppler Effect and Racing Cars'', Physics
    Teacher 26, 423 (1988). 
  \bibitem{Saba01}
    M. M. F. Saba and R. A. da S. Rosa, ``A quantitative demonstration
    of the Doppler effect'', Phys. Teach. 39,
    431--433 (2001).
  \bibitem{Octave}\url{http://www.gnu.org/software/octave/}
  \bibitem{sox}\url{http://sox.sourceforge.net/}
  \bibitem{Fresnel}M. Abramowitz, I.A. Stegun, Handbook of
    Mathematical Functions, Dover Publications, New York (1972). p. 295.
\end{thebibliography}
\end{document}